\documentclass{article}
\usepackage{amsmath}
\usepackage{authblk}
\usepackage{booktabs}
\usepackage[margin=2cm]{geometry}
\usepackage[dvipsnames,table]{xcolor}
\usepackage[colorlinks=True, linkcolor=MidnightBlue, citecolor=ForestGreen]{hyperref}
\usepackage{siunitx}

\title{Two be or not two be: the question of \\ flux conventions in neutrino astronomy}
\author[1]{Mathieu Lamoureux}
\affil[1]{Universit\'e Paris Cit\'e, CNRS, Astroparticule et Cosmologie, F-75013 Paris, France}

\begin{document}

\maketitle

\begin{abstract}
    When reporting astrophysical neutrino fluxes and detector performance, it is common to consider the contributions from both neutrinos and antineutrinos, and from the three flavours ($e$, $\mu$, $\tau$). However, this can lead to missing factors of $2-3-6$ in the interpretation, especially because it can be unclear whether the flux is averaged between or summed over these components. In addition, other factors may come into play depending on what is reported, especially when flux, energy flux, and fluence are used simultaneously. This article presents the related issues and how they may be mitigated by being more specific in the notation and presentation of the results.
\end{abstract}

\section{Introduction}

Compared to other cosmic messengers, neutrinos have the peculiarity of coming in different flavours. Hence, as opposed to photons, for which a single flux can be provided to describe the observations fully, astrophysical neutrinos would in principle require six fluxes, one per chirality ($c = \nu$, $\bar\nu$) and per flavour ($\alpha = e, \mu, \tau$). The differential neutrino flux (flux in the following) can be written as
\begin{equation}
    \dfrac{{\rm d}^6 N}{{\rm d}E\,{\rm d}t\,{\rm d}^2A\,{\rm d}^2\Omega} \equiv \Phi_{\rm total} = \sum_{c \in \{\nu, \bar\nu\}} \sum_{\alpha \in \{e, \mu, \tau\}} \Phi_{c, \alpha},
    \label{eq:flux}
\end{equation}
where the different derivatives indicate that the flux is given per unit of energy, time, surface, and solid angle (in units of \unit{\per\GeV\per\s\per\square\cm\per\steradian}).

In most cases, the dominant astrophysical neutrino production mechanism is via cosmic-ray interactions with the medium, producing charged pions which further decay into muons, leading to a ratio at the source of about $\nu_e:\nu_\mu:\nu_\tau = 1:2:0$. After oscillations over astrophysical distances, this translates into a ratio at Earth of about $1:1:1$, which is referred to as ``equipartition''~\cite{Learned:1994wg}. An equal split between neutrinos and antineutrinos is also assumed. In these conditions, \autoref{eq:flux} may be simplified to
\begin{equation}
    \Phi_{\rm total} = 3 \times \Phi_{\nu+\bar\nu}^{\rm per-flavour} = 6 \times \Phi_0,
    \label{eq:flux-simple}
\end{equation}
where $\Phi_{\nu+\bar\nu}^{\rm per-flavour}$ is the $\nu+\bar\nu$ per-flavour flux, and $\Phi_0 = \Phi_{c,\alpha}$ (for any $c$ and $\alpha$) is the per-flavour and per-chirality flux (in the same units as above). The conditions of equipartition may be relaxed, for instance when considering processes beyond the standard production scenario (changing the ratio at the source) or effects during the propagation (for instance, Lorentz Invariance Violation), but these cases are beyond the scope of this article.

Though the raised points may appear trivial and straightforward, they are subject to numerous internal discussions and checks, as well as several errata in the literature. These two aspects have been the motivations for summarising the problem and how it could be addressed in this article.

In the following, two main points will be addressed: \autoref{sec:2} describes how \autoref{eq:flux-simple} may lead to missing scaling factors in the interpretation of the results, e.g., when deriving upper limits on the flux normalisation; confusion between different flux or fluence definitions may also induce troubles when comparing measurements in the literature, as detailed in \autoref{sec:other}.

\section{The ``factor two'' problem}
\label{sec:2}

The expected number of signal events in a given neutrino telescope is directly related to the neutrino flux:
\begin{equation}
    N_{\rm sig} = \int A_{\rm eff}(E, \Omega, t) \times \Phi(E, \Omega, t)\,{\rm d}E\,{\rm d}\Omega\,{\rm d}t \equiv A_{\rm eff} \otimes \Phi
    \label{eq:nsig}
\end{equation}
where $A_{\rm eff}(E, \Omega, t) = \sum_{c,\alpha} A_{\rm eff,c,\alpha}$ is the all-flavour time-dependent neutrino effective area. The problem occurs when considering which fluxes and effective areas should be used in \autoref{eq:nsig}. Let's decompose the problem per flavour and chirality:
\begin{align}
    N_{\rm sig} &= \sum_{c,\alpha} N_{\rm sig,c,\alpha} \\
                &= \sum_{c,\alpha} A_{\rm eff,c,\alpha} \otimes \Phi_{c, \alpha} \nonumber \\
                &= \sum_{c,\alpha} A_{\rm eff,c,\alpha} \otimes \Phi_0 \nonumber \\
                &= A_{\rm eff,total} \otimes \Phi_0
\end{align}
where $A_{\rm eff,total} = \sum_{c,\alpha} A_{\rm eff,c,\alpha}$ is the total effective area, and the last two lines are valid only in the equipartition hypothesis. Therefore, the total number of signal is obtained by combining the effective area, \emph{summed} over all flavours and chirality, with the flux \emph{per} flavour and chirality. A common mistake is to use a different flux convention. For instance, articles often report the per-flavour $\nu+\bar\nu$ flux, $\Phi_{\nu+\bar\nu}^{\rm per-flavour}$, which would give an additional factor of $2$ in $N_{\rm sig}$ when combined with the total effective area.

When converting observations to a constraint on the flux, e.g., a best fit or an upper limit, this may go completely unnoticed as this is a simple scaling. For instance, if a per-flavour per-chirality flux $\Phi_0(E; \phi) = \phi \times f(E)$ is assumed, the statistical analysis provides an upper limit on the number of signal events $N_{\rm sig} < N_{\rm sig}^{\rm up}$ and the corresponding upper limit on the flux normalisation $\phi$ is obtained as 
\begin{equation}
    \phi^{\rm up} = \phi^{\rm ref} \times \dfrac{N_{\rm sig}^{\rm up}}{N_{\rm sig}(\phi^{\rm ref})},
\end{equation}
where $\phi^{\rm ref}$ is a reference flux for which the expected number of signal events is known. If a factor $\alpha$ slips in the computation of $N_{\rm sig}(\phi^{\rm ref})$, the flux limit is then wrong by a factor $\alpha^{-1}$.

\autoref{tab:corr-factor} presents the correction factors that are needed to get the flux constraint in the same convention as the reference flux. As emission models are usually provided in terms of the per-flavour $\nu+\bar\nu$ flux, it is often convenient to compute and use the effective area $A_{\rm eff,(\nu+\bar\nu)/2}^{\rm all-flavour} = A_{\rm eff, total} / 2$.

\begin{table}[hbtp]
    \centering
    \begin{tabular}{c|ccc}
        \toprule
        $\Phi$ \textbackslash $A_{\rm eff}$ & $A_{\rm eff,c,\alpha}$~$^a$ & $A_{\rm eff,(\nu+\bar\nu)/2}^{\rm all-flavour}$~$^b$ & $A_{\rm eff,total}$ \\
        \midrule
        $\Phi_0$ & $1/6$ & $1/2$ & \cellcolor{green!15}$1$ \\
        $\Phi_{\nu+\bar\nu}^{\rm per-flavour}$ & $1/2$ & \cellcolor{green!15}$1$ & $3$ \\
        $\Phi_{\rm total}$ & \cellcolor{green!15}$1$ & $3$ & $6$ \\
        \bottomrule
    \end{tabular}
    \caption{Factor by which the flux constraint should be multiplied to get a correct result for the flux, if a wrong combination is used, assuming equipartition between flavours and chiralities. The green-coloured cells show the cases where no correction is needed. \\[0.3em] \scriptsize a: The correction in this column applies only if the detector response is the same for all the flavours and chiralities. \\ \scriptsize b: The correction in this column applies only if the detector response is the same for neutrinos and antineutrinos.}
    \label{tab:corr-factor}
\end{table}

However, it may not be obvious whether the factors $2$ and $3$ are already accounted for in the generation weight provided by the simulation software or corrected when combining files for different interaction channels. It is therefore crucial to make all the notations explicit and document precisely how these weights should be used. Additionally, notations in the literature are sometimes not explicit about the convention used when plotting effective area or reporting upper limits. 

The recommendation is to \emph{always} report the convention used in text, figures, and tables, and to include whether assumptions are made (e.g., equipartition) to account for different flavours and chiralities.

\section{Different flux conventions}
\label{sec:other}

Other common mistakes occur when different notions related to flux are entangled. So far, we have been referring to the ``differential neutrino flux'' $\Phi$, defined as the number of neutrinos crossing a unit surface (${\rm d}A$), from a unit solid angle (${\rm d}\Omega$), in a given unit time (${\rm d}t$) and within a given energy slice (${\rm d}E$). This quantity is often directly relevant when considering a diffuse emission, either isotropic over the whole sky ($\Phi = \Phi(E[, t])$) or following some template ($\Phi = \Phi(E, \Omega[, t])$), e.g., in the case of the emission from the Galactic Plane.

However, when considering the emission from specific sources, other quantities may be more relevant, introducing more potential sources of mistakes, as discussed in the following subsections.

\subsection{Notations}

The following paragraphs introduce notations commonly used in the literature.

\paragraph{Point-source flux}
In the case of a single point-like source, the direction dependence is irrelevant, and we may write the differential flux as
\begin{equation}
    \Phi^{\rm PS} = \dfrac{{\rm d}^4 N}{{\rm d}E\,{\rm d}t\,{\rm d}^2A},
\end{equation}
in units of \unit{\per\GeV\per\square\cm\per\s}. Even in the case of diffuse emission, we may integrate over the solid angle to get a differential flux in the same units. \textit{In the following, we will therefore drop the solid angle from the flux ($\Phi \equiv \Phi_{\rm PS}$) without loss of generality, as it can be reintroduced at any point in the notation.}

\paragraph{Energy-squared flux}
As the flux typically follows a power law $E^{-\Gamma}$ with $\Gamma \in [2,3]$, it is often convenient to rather represent it as $E^2 \Phi$, in units of \unit{\GeV\per\square\cm\per\s}, instead of $\Phi$ directly, allowing one to flatten the main slope of the spectrum and highlight the deviations from an $E^{-2}$ power law. In the simple case where a spectrum $\propto E^{-2}$ is assumed, the quantity is energy-independent, allowing easy comparison between experiments.

\paragraph{Particle or energy flux}
In some cases, we may not care about the energy distribution of events, but rather about the number or the amount of energy crossing a unit surface. The particle flux $F_P$ and energy flux $F_E$ are defined as
\begin{align}
    F_P(t) &= \int_{E_{\min}}^{E_{\max}} \Phi(E, t) \, {\rm d}E, \\
    F_E(t) &= \int_{E_{\min}}^{E_{\max}} E \times \Phi(E, t) \, {\rm d}E,
\end{align}
in units of \unit{\per\square\cm\per\s} and \unit{\GeV\per\square\cm\per\s}, respectively. These quantities depend on the chosen integration bounds $[E_{\rm min}, E_{\rm max}]$, which should be explicitly specified when reporting these values, especially when one wants to compare with the constraints from another article.

\paragraph{Fluence}
Going one step further, we can integrate the energy flux over time to compute the fluence
\begin{equation}
    \mathcal{F} = \int E \times \Phi(E, t) \, {\rm d}E {\rm d}t,
    \label{eq:fluence}
\end{equation}
in units of \unit{\GeV\per\square\cm}. It represents the total amount of energy that has crossed a unit surface from the considered emission mechanism. Both the energy flux and the fluence can be used to estimate the source luminosity and the total energy emitted by the source, after correcting for the source distance and the energy shift.

\subsection{Possible mistakes}

Though energy flux and energy-squared flux have the same units, they are different quantities. For instance, assuming a single power law $E^{-\Gamma}$, we get:
\begin{align}
    E^2 \Phi(E, t) &= \phi E^{-\Gamma+2} \\
    F_E(t) &= \left\{ \begin{array}{ll}
        \phi \times \log \dfrac{E_{\max}}{E_{\min}} & \text{for } \Gamma = 2 \\[1em]
        \phi \times \dfrac{E_{\max}^{-\Gamma+2} - E_{\min}^{-\Gamma+2}}{-\Gamma + 2} & \text{for } \Gamma \neq 2
        \end{array} \right.
\end{align}
For the usual $\Gamma = 2$ case, $E^2 \Phi$ and $F_E$ are the same up to a constant term $\log(E_{\max}/E_{\min})$, complicating investigations when comparing different results and trying to figure out which ``flux'' convention has been used.

Similarly, in the case of transient sources, it is often convenient to quote the time-integrated energy-squared flux $\int E^2 \Phi(E, t) {\rm d}t$, which has the same units as the fluence in \autoref{eq:fluence}. It is sometimes incorrectly referred to as ``fluence'', resulting in misleading comparisons.

The recommendation to the community would be to take the habit of describing precisely what is defined as ``flux'', always starting from the notation $\dfrac{{\rm d}^6 N}{{\rm d}E\,{\rm d}t\,{\rm d}^2A\,{\rm d}^2\Omega}$ or $\dfrac{{\rm d}^4 N}{{\rm d}E\,{\rm d}t\,{\rm d}^2A}$ and deriving useful quantities from there.

\section{Conclusion}

We discussed several common sources of confusion arising from the peculiarity of neutrinos coming in three flavours and two chiralities, as well as from the definition of the ``flux'' quantity. The list is not exhaustive, as there are always other factors ($=2$ or $\neq 2$) that may hide in the details of an analysis, such as solid angle averaging (when computing all-sky or upgoing sky effective areas), or use of natural vs base-10 logarithm.

We recommend explicitly stating the conventions used and defining all reported quantities. Collaborations also have a strong incentive to ensure consistency across publications, so that results can be compared with previous findings and other experiments. Public datasets should also be clearly labelled to avoid potential misuse.

\section*{Acknowledgements}

M.L. has received funding from the European Union’s Horizon Europe Research and Innovation programme under Grant Agreement No 101131928.

\end{document}